\documentclass[usenatbib]{mnras}

\usepackage{newtxtext,newtxmath}

\usepackage[T1]{fontenc}
\usepackage{ae,aecompl}
\usepackage{tabularx}

\DeclareRobustCommand{\VAN}[3]{#2}
\let\VANthebibliography\thebibliography
\def\thebibliography{\DeclareRobustCommand{\VAN}[3]{##3}\VANthebibliography}

\usepackage{graphicx}
\usepackage{amsmath}
\usepackage[l3]{csvsimple}
\usepackage{siunitx,array,booktabs,xfp}
\usepackage{orcidlink}

\newcommand{\Msun}{${\rm M}_{\odot}$ }

\title[Mock Imaging vs. Intrinsic Properties]{Size-Mass Relations for Simulated Low-Mass Galaxies: Mock Imaging versus Intrinsic Properties}

\author[C. Klein et al.]{\hspace{-.01cm}Courtney Klein$^{1}$\thanks{E-mail: court.klein32@gmail.com} \orcidlink{0000-0002-2762-4046}, 
James S. Bullock$^{1}$ \orcidlink{0000-0003-4298-5082},
Jorge Moreno$^{2}$$^{,}$$^{3}$ \orcidlink{0000-0002-3430-3232},
Francisco J. Mercado$^{2}$$^{,}$$^{4}$ \orcidlink{0000-0002-5908-737X}, 
\newauthor
Philip F. Hopkins$^{4}$ \orcidlink{0000-0003-3729-1684},
Rachel K. Cochrane$^{5}$$^{,}$$^{6}$ \orcidlink{0000-0001-8855-6107},
Jose A. Benavides$^{7}$ \orcidlink{0000-0003-1896-0424}
\vspace*{5pt} \\
\\
$^{1}$Department of Physics and Astronomy, University of California Irvine, Irvine, CA 92697, USA \\
$^{2}$Department of Physics and Astronomy, Pomona College, Claremont, CA 91711, USA \\
$^{3}$The Observatories of the Carnegie Institution for Science, 813 Santa Barbara Street, Pasadena, CA 91101, USA \\
$^{4}$TAPIR, Mailcode 350-17, California Institute of Technology, Pasadena, CA 91125, USA\\
$^{5}$Department of Astronomy, Columbia University, New York, NY 10027, USA; \\
$^{6}$Center for Computational Astrophysics, Flatiron Institute, 162 Fifth Avenue, New York, NY 10010, USA \\
$^{7}$Department of Physics and Astronomy, University of California, Riverside, 900 University Avenue, Riverside, CA 92521, USA}

\date{Accepted 2024 June 11. Received 2024 May 28; in original form 2024 April 04}

\pubyear{2024}

\begin{document}
\label{firstpage}
\pagerange{\pageref{firstpage}--\pageref{lastpage}}
\maketitle

\begin{abstract}
The observationally-inferred size versus stellar-mass relationship (SMR) for low-mass galaxies provides an important test for galaxy formation models. However, the relationship relies on assumptions that relate observed luminosity profiles to underlying stellar mass profiles. Here we use the Feedback in Realistic Environments simulations of low-mass galaxies to explore how the predicted SMR changes depending on whether one uses star-particle counts directly or mock observations. We reproduce the SMR found in The Exploration of Local Volume Satellites survey remarkably well only when we infer stellar masses and sizes using mock observations. However, when we use star particles to directly infer stellar masses and half-mass radii, we find that our galaxies are too large and obey a SMR with too little scatter compared to observations. This discrepancy between the ``true'' galaxy size and mass and those derived in the mock observation approach is twofold. First, our simulated galaxies have higher and more varied MLRs at a fixed colour than those commonly-adopted, which tends to underestimate their stellar masses compared to their true, simulated values. Second, our galaxies have radially increasing MLR gradients therefore using a single MLR tends to under-predict the mass in the outer regions. Similarly, the true half-mass radius is larger than the half-light radius because the light is more concentrated than the mass. If our simulations are accurate representations of the real universe, then the relationship between galaxy size and stellar mass is even tighter for low-mass galaxies than is commonly inferred from observed relations.
\end{abstract}

\begin{keywords}
galaxies: dwarf -- galaxies: fundamental parameters -- galaxies: evolution
\end{keywords}



\section{Introduction}

Our ability to accurately compare predictions from galaxy formation simulations to observations is fundamental in testing and improving physical understanding. Among the most important physics we want to test include fundamental dark matter physics and the emergent physics of stellar feedback; dwarf galaxies are ideal laboratories for both. 
These galaxies tend to be dark matter dominated and strongly affected by stellar feedback processes owing to their shallow potential wells, both of which will drive galaxy morphology \citep[e.g.][]{BBK17}.  

The observed size-mass relationship (SMR) of galaxies provides an important means to test models and thereby probe galaxy formation and evolution physics \citep{Rodriguez2021,Huang2017,Shen2003TheSurvey}. For example, in rotationally-supported galaxies, size is linked to angular momentum conservation, with ties to the spin distribution in dark matter halos \citep[e.g.][]{Peebles69,Mo2010}. In dispersion-supported galaxies, sizes are more directly linked to turbulence in the star-forming interstellar medium or energy exchange associated with mergers \citep{MBK2006,Kaufmann2007OnGalaxies,Trujillo2011}. For dispersion-supported dwarf galaxies, in particular, feedback processes are likely important in setting galaxy sizes \citep[e.g.][]{El-Badry16}. 

The SMR of dwarf galaxies has been the focus of several observational surveys, including TiNy Titans \citep{Stierwalt2015TiNyEvolution} and Exploration of Local Volume Satellites survey (ELVES) \citep{Carlsten2021StructuresShapes}. Larger volume surveys have explored a wide range of sizes and stages of galaxy evolution. Work using data from HST CANDELS \citep{vanderWel20143D-HST+CANDELS:3} and Cosmic Evolution Survey (COSMOS)/Drift And SHift (DASH) \citep{Mowla2019COSMOS-DASH:iHST/i} has focused on the SMR of galaxies up to a redshift of 3 but is limited to stellar masses greater than 3x10$^9$$M_{\odot}$.  \cite{George2024TwoNewcomers} identifies the impact of bulge evolution and recently quenched galaxies on the SMR of galaxies with stellar masses greater than 10$^{9.5}$$M_{\odot}$ up to a redshift of 1. The Galaxy and Mass Assembly (GAMA) Survey uses the Anglo-Australian Telescope (AAT) to build off of Sloan Digital Sky Survey (SDSS) galaxy observations to create a large database of low redshift galaxies \citep{Driver2011GalaxyRelease, Liske2015Galaxy2}. \cite{Lange2015GalaxyMorphology, Lange2016GalaxySpheroids} use GAMA to explore the SMR of galaxies based on different galaxy types and components respectively.

\begin{figure*}
  \centering
\includegraphics[width=\textwidth]{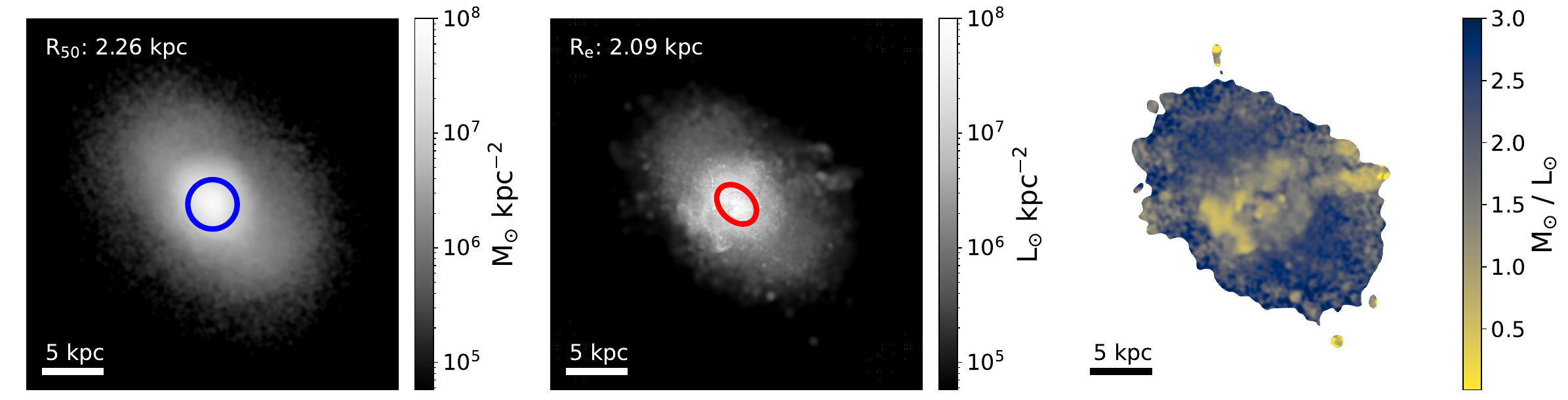}
  \caption{\textit{Left panel:} A particle-based stellar mass surface density rendering of simulated galaxy mllc ($M_{\star} \sim 10^9$M$_\odot$) viewed in the ``xy'' orientation. The blue circle has a radius equal to the particle-derived 2D half-mass radius R$_{50}$. \textit{Middle panel:} A mock $g$-band image of the same galaxy viewed along the same orientation. The red ellipse has an axis ratio, orientation, and effective radius $R_e$ along the major axis found through a S\'ersic profile fit. \textit{Right panel:} The implied MLR map for this galaxy. Note that the MLR varies significantly over the surface of the galaxy and is generally higher at large radii.}
  \label{fig:part_mock_circle}
\end{figure*}

When exploring the SMR it is critical to understand how galaxy size and mass are defined. The most commonly-adopted radii used in observations is the effective radius ($R_e$), which is the radius that contains half of the galaxy's light when assuming an underlying S\'ersic surface brightness profile. A two-dimensional S\'ersic profile is often used with fixed ellipticity, such that the effective radius is defined along the galaxy's major axis \citep{Shen2003TheSurvey, vanderWel2012STRUCTURALCANDELS, Stierwalt2015TiNyEvolution, Carlsten2021StructuresShapes, Wang2023UNCOVER:Galaxies}. The Petrosian radius \citep{Petrosian1976SurfaceGalaxies} is another common choice, adopted by integral field spectrograph surveys such as Sloan Digital Sky Survey \citep{Shen2003TheSurvey} in their Mapping Nearby Galaxies at Apache Point Observatory (MaNGA) program \citep{Blanton2017SloanUniverse}. Another option is to chose the radius corresponding to a fixed surface brightness or derived mass surface density. Some popular choices include 1 $M_{\odot}$ pc$^{-2}$ \citep{Trujillo2020AImaging}, $\mu_B$ = 26.5 mag arcsec$^{-2}$ \citep{Holmberg1958ANebulae.}, and $\mu_g$ = 23.5 mag arcsec$^{-2}$ which has been seen to reduce the radial scatter in the galaxy size-luminosity relation \citep{Hall2012AnGalaxies}.

The mass of a galaxy must be estimated observationally using colours or spectra to infer the stellar populations. Spectral energy distribution fitting is the gold standard, but even this method is model-dependent \citep{Lower2020HowFitting}. When a galaxy spectra is not available, broad-band photometric data may be used to infer a model-dependent stellar-mass-to-light ratio \citep[e.g.][]{Taylor2011GalaxyEstimates,Into2013NewDust}. 

Theorists have been utilizing the SMR to test physics implemented in simulations \citep{Mo98}, with numerical simulations playing an increasingly important role in these interpretations \citep[e.g.][]{SGK18,Jiang19,Rohr2022TheFIRE}. Often, different simulation comparisons adopt varied definitions of galaxy size and mass. For example, \citet{Arora2023MaNGARelations} use the Numerical Investigation of a Hundred Astrophysical Objects (NIHAO) simulations and define radius using a characteristic stellar mass surface density $\Sigma_*$ = 10 \Msun pc$^{-2}$ for both observations in MaNGA and in NIHAO. \citet{DeGraaff2022ObservedObservations} uses S\'ersic profiles fit to both mock light and mass images developed from the Evolution and Assembly of GaLaxies and their Environments (EAGLE) simulations. They find good agreement with the observed SMR from GAMA for galaxies with stellar masses larger than $10^{10}$ \Msun and between masses derived from S\'ersic profiles fit to mock mass images and mass enclosed in 30 kpc. In another example, \cite{Benavides2023OriginEnvironments} uses IllustrisTNG simulations to understand the origin of Ultra Diffuse Galaxies using the half-mass radius of the stars as a measure of galaxy size and explores the driving factors that cause these outliers in the SMR. Massive FIRE galaxies between z $\sim$ 1.25 and z $\sim$ 2.75 were studied using mock images to explore the impact of stellar feedback and AGN feedback on galaxy sizes \citep{Cochrane2023TheSizes,Parsotan2020RealisticSimulations}. The SMR has the power to test our simulations abilities to reproduce observed galaxies and so we explore the low-mass regime of FIRE along with different methods of measuring the galaxies.

In this paper, we explore the difference between two distinct methods for determining galaxy size and stellar mass within our simulations. Specifically, we look at a Particle Method and a Mock Observation Method. We are motivated to compare what is common among simulators -- using star particles directly to infer half-mass radii and stellar masses of simulated galaxies -- to a method that more directly follows what is often done in observational survey papers. One question we wish to ask is how well observables trace the underlying physical properties of simulated galaxies and what biases are present when comparing the two. Another question is whether mock observations allow a better match to observed trends compared to a more simplified particle-based approach. We present our simulation sample along with our Particle and Mock Observation Methods for measuring our galaxies in Section \ref{sec:methods}. In Section \ref{sec:results} we use the SMR to compare the two methods to each other and to an observational sample. We summarize our results and discuss implications in Section \ref{sec:conclusions}.

\section{Methods}
\label{sec:methods}

\begin{table*}
\caption{Simulated galaxies (initial gas mass resolution) and their measured properties for three orientations (``View''). $M_{\star}^{\rm part}$ is the particle-derived stellar mass and $R_{50}$ is the particle-derived stellar half-mass radius. $M_{\star}^{\rm mock}$ is the stellar mass inferred from our mock-observation method. $R_e$, $n$, and $b/a$ are from S\'ersic profile fits. $L_g$ is the S\'erisc integrated $g$-band luminosity, and $g-r$ are the associated colours.  $M/L$ is the``true'' simulated MLR, $M_{\star}^{\rm part}/L_{\rm image}$.}
\vspace{-0.5\baselineskip} 
\label{tab:sims}

  \csvreader[
    head to column names, 
    before reading = \begin{center},
    tabular=lcccccccccc,
    table head = \toprule \textbf{Simulation} & \textbf{View}  & \textbf{$M_{\star}^{\rm part}$} [$M_{\odot}$] & \textbf{$R_{50}$} [kpc]& \textbf{$M_{\star}^{\rm mock}$} [$M_{\odot}$] & \textbf{$R_{e}$}[kpc]& \textbf{$n$} & \textbf{$b/a$} & \textbf{$L_g$} [L$_{\odot}$] & \textbf{$g-r$}& \textbf{$M/L$}\\\midrule,
    late after line=\\,
    table foot = \bottomrule,
    after reading = \end{center}]
    {data/datapaper.csv}{}
    {\sim & 
    \view & 
    \tablenum[exponent-mode=scientific, round-precision=2, round-mode=places]{\mpart} & 
    \tablenum[round-precision=2,round-mode=places]{\rpart} &
    \tablenum[exponent-mode=scientific, round-precision=2, round-mode=places]{\mmock} & 
    \tablenum[round-precision=2,round-mode=places]{\rmock} &
    \tablenum[round-precision=2,round-mode=places]{\n} &
    \tablenum[round-precision=2,round-mode=places]{\ba} &
    \tablenum[exponent-mode=scientific, round-precision=2, round-mode=places]{\lg} &
    \tablenum[round-precision=2,round-mode=places]{\gr} &
    \tablenum[round-precision=2,round-mode=places]{\ml}
    }

\end{table*}

\subsection{Simulations}
We analyze 20 zoom-in simulations of low-mass isolated galaxies from the Feedback In Realistic Environments (FIRE) Project\footnote{\url{https://fire.northwestern.edu/}}, version FIRE-2 \citep{Hopkins2018FIRE-2Formation}. It was run using \texttt{GIZMO}\footnote{\url{http://www.tapir.caltech.edu/~phopkins/Site/GIZMO.html}} with a meshless finite-mass Lagrangian Godunov method which modifies spatial resolution to maintain constant mass elements and conserves mass, momentum, and energy. \texttt{GIZMO} solves for the hydrodynamics and gravity in a fully adaptive Lagrangian method. 

These simulations include feedback from Type Ia and II supernovae, stellar mass loss from OB and AGB stars, and radiative processes such as radiation pressure, photo-electric effect, and photo-ionization heating. They track gas heating and cooling for temperatures 10 - 10$^{10}$ K along with 11 stellar and gas elemental abundances: H, He, C, N, O, Ne, Mg, Si, S, Ca, and Fe. The criteria for star formation is molecular, self-gravitating, and self-shielding gas with a minimum density of 1000 cm$^{-3}$ that is Jeans Unstable. We assume a flat $\Lambda$CDM cosmology with $\Omega_{m}$ $\approx$ 0.32, $\Omega_{\lambda} = 1 - \Omega_{m}$, $\Omega_{b}$ $\approx$ 0.049, and $H_0$ $\approx$ 67 km s$^{-1}$ Mpc$^{-1}$ \citep{PlankCollaboration2014PlanckParameters}.

The sample we use consists of galaxies with dark matter halo masses of 10$^{10}$ ${\rm M}_{\odot}$ (m10's) and 10$^{11}$ ${\rm M}_{\odot}$ (m11's) at z=0. The m10s/m11s have baryonic mass resolution of 500/2100 M$_{\odot}$ and a force resolution of 2 pc \citep{Fitts17}. The initial conditions for these halos were chosen in accordance with the methods discussed in \cite{Onorbe2014HowSimulations}, and in that respect should represent a fair sample of halo properties across the mass range. A summary of these simulations and their properties is provided in Table \ref{tab:sims}.

\begin{figure*}
  \centering
\includegraphics[width=.9\textwidth]{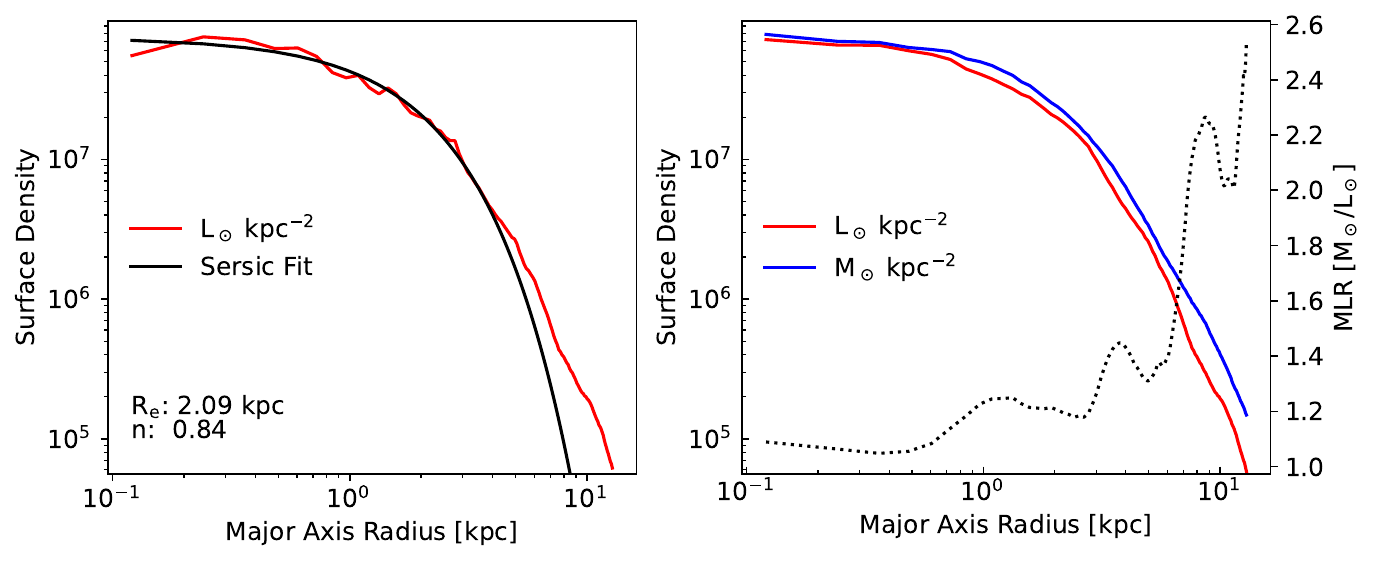}
  \caption{\textit{Left panel:} Luminosity surface brightness for galaxy m11c. The red line shows the elliptical azimuthally-averaged surface brightness profile as a function of major axis radius and the black line is the best-fit S\'ersic profile. The reported $R_e$ and $n$ are the best fit values. \textit{Right panel:} Average surface brightness and surface mass density along the major axis in red and blue, respectively. The dotted black line (right vertical axis of the plot) shows the average MLR along the major axis.}
  \label{fig:sersic}
\end{figure*}

\subsection{Galaxy Properties: Particle Method}
\label{sec:partmethod}

To measure the ``true'' size and mass of our simulated galaxies we use the star particles directly. We only consider stellar particles within 10\% of the dark matter virial radius, 0.1 $R_{\rm vir}$. We visually inspect each galaxy to ensure there are no satellite galaxies included within this region. The total ``particle-method'' stellar mass reported for each galaxy, $M_{\star}^{\rm part}$, is the summed mass of these stellar particles. Note that we define all virial quantities using the \citet{BN98} definition of the virial overdensity. As an example, with this choice of virial overdensity, $M_{\rm vir} = 10^{11}$ M$_\odot$ corresponds to $R_{\rm vir} \approx 130$ kpc at $z=0$. 

To define the half-mass radius with this method, we first project the stellar mass in a random orientation. One such projection is demonstrated in the stellar mass surface density image in the left panel of Figure \ref{fig:part_mock_circle} for galaxy m11c. Note that the smoothed mock mass image is only used for demonstrative purposes and the particle derived measurements assume point sources. We define the particle-based half-mass radius, $R_{50}$, as the radius of a circle that encloses half of the (particle-based) stellar mass. The blue circle in the left panel of Figure \ref{fig:part_mock_circle} shows $R_{50}$ for this galaxy projection, as an example. 

Note that while we treat $R_{50}$ as a ``true'' or intrinsic quantity, its value typically depends on the projected orientation of the simulated galaxy since we are assuming circular symmetry for 2D projections of systems that are not spherically symmetric. We adopt this approach due to its common use in the theoretical literature and to more accurately compare with observation which are 2D projections. 

In what follows, we view each galaxy three times, one each for the default $x$-$y$-$z$ axes of the simulation box.  We then measure particle-derived properties (as described above) and mock-derived properties (as described below) for each projection.  These are random orientations from a physical standpoint. The properties derived for all three projections for each galaxy are listed in Table \ref{tab:sims}. Note that $R_{50}$, in particular, changes noticeably with orientation, often at the $\sim 20-30\%$ level.

\subsection{Galaxy Properties: Mock Observation Method}
\label{sec:mockobs}

We create mock images for our simulated galaxies using \texttt{FIRE\textunderscore studio} \citep{Gurvich2022FIRESimulations, Hopkins2005ALifetimes} with the goal of producing images limited solely by the simulation resolution and not other implemented observational limitations. To determine the luminosity of each star particle we use its age and stellar mass along with a catalog of MLRs derived from evolving simple stellar populations with a Chabrier Initial Mass Function and a variety of metallicities \citep{Bruzual2003Stellar2003}. Using the metallicity and age of the stellar particle, we use the corresponding MLR in our desired observing band and then scale the luminosity with the particle's mass. In what follows we use SDSS $g$ and $r$ bands.

We take extinction into account via Thomson scattering, the photoelectric effect, and dust extinction \citep{Hopkins2005ALifetimes}. We use a standard Thomson scattering cross section and the photoelectric effect absorption cross section derived in \cite{Morrison1983InterstellarKeV} for our given wavelength band of interest. We use the dust scattering cross section and extinction curves from \cite{Pei1992InterstellarClouds} for the Small Magellanic Cloud. These cross sections are linearly combined with the dust scattering cross section scaled by a dust-to-gas ratio of 0.78 \citep{Pei1992InterstellarClouds}. The cross section is derived for solar metallicity and is later scaled by the metallicity of a given gas particle.

Both the star and gas particles are then smoothed out using a cubic spline kernel with a smoothing length twice that of the simulation particle smoothing length. We use a $1000\,{\times}1000$ pixel image, such that the pixel size is less than the median smoothing length to ensure image resolution is limited by the simulation and not the pixel size. The field of view (FOV) for each image is initially based on 0.1 $R_{\rm vir}$, however the FOV is decreased if the galaxy only occupies less than half of the image. Our pixel sizes range from 6-80 pc, depending on the galaxy. Note that our approach is to construct simulation-limited ``idealized'' images. In this work we ignore additional observational limitations associated with sky background or instrument resolution, though it would be interesting to explore the impact of these effects on the radii and masses measured.

For each orientation, the ray projection proceeds from back to front. We integrate the light along the line of sight using any portion of the light from the smoothed star particles that falls within the relevant pixel. When the ray projection encounters a smoothed gas particle, it damps the integrated luminosity by e$^{-\kappa*\mu}$, where $\kappa$ is the opacity derived by scaling the cross section with the average scattering particles (not simulation particles) per unit mass. The quantity $\mu$ is the surface mass of gas with temperatures less than $10^5$K within the pixel, scaled by its gas metallicity. We do not account for light scattered into the line-of-sight.

\begin{figure*}
  \centering  \includegraphics[width=.9\textwidth]{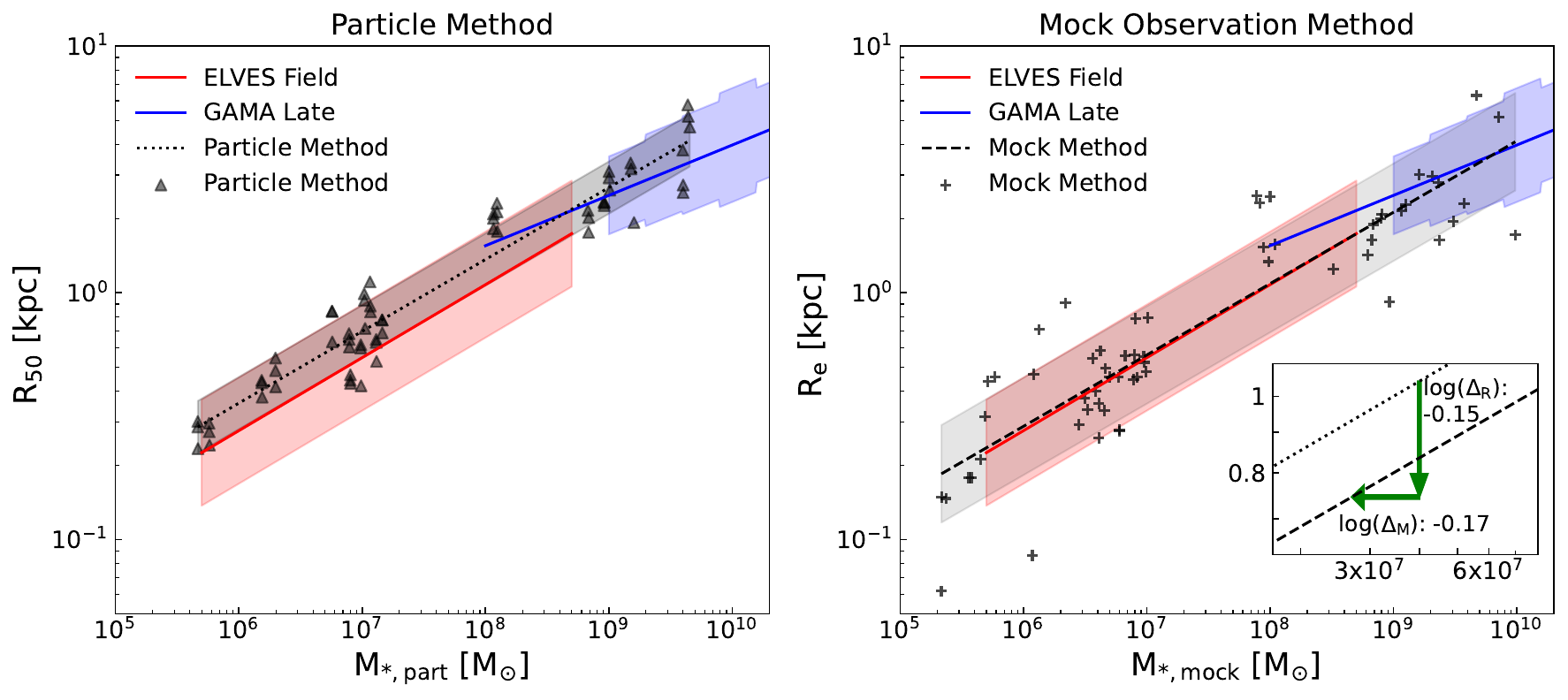}
  \caption{
  \textit{Left panel:} SMR for the simulated galaxies using the particle method. The grey triangles represent the individual galaxies. The black dotted line is the best fit relation to the galaxies with the grey shaded region representing the 1$\sigma$ scatter in the simulation data. The red line and shading represents the ELVES field galaxies and their 1$\sigma$ scatter. The blue line and shaded region represents the GAMA late type galaxies and their 1$\sigma$ scatter. 
  \textit{Right panel:} Same as the left panel, except the grey points are now determined from our mock method and the black dashed line and shaded bands are best fits and 1$\sigma$ ranges for those points. The zoomed in plot on the bottom right shows the particle method relation as a dotted line and the mock method relation as a dashed line. The down-pointing green arrow indicated the average scaling factor between R$_e$ and R$_{50}$. The left-pointing green arrow shows the average scaling factor between $M_{\star}^{\rm mock}$ and $M_{\star}^{\rm part}$.
  }
  \label{fig:mvr}
\end{figure*}

Using the mock images, we determine the galaxy's radius by fitting a S\'ersic profile to an elliptical azimuthally-averaged radial surface brightness profile \citep{Sersic1963InfluenceGalaxy}. We first use Astropy module \texttt{Sersic2D} solely to determine the orientation $\theta$ and ellipticity $e = 1 - b/a$ of the galaxy on the sky \citep{Astropy2022}. Here, $a$ and $b$ are the lengths of the semi-major and semi-minor axes of each elliptical isophote. We then construct an averaged radial surface brightness profile using elliptical annuli of fixed ellipticity sampled 100 times linearly from the center to the FOV along the major axis. This surface brightness profile is then fit with 1D S\'ersic profile as a function of major axis $r$:

\begin{equation}
  I(r) = I_e \exp \bigg\{ -b_n \bigg[ \Big( \frac{r}{R_e}\Big) ^{1/n}- 1 \bigg] \bigg\}.
\end{equation}

Here, $R_e$ is the effective half-light radius along the major axis, $I_e$ is the intensity at $R_e$, and $n$ is the S\'ersic index. Note that if there are fewer than 15 radial bins within $R_e$ then the sample number is iteratively increased. The center region of the galaxy strongly effects the fit and therefore we ensure it is sufficiently sampled. We found that requiring more than 20 samples began to introduce noise in the profile as the bin size approached the pixel size. We also cut the fit off at the 30 $g$-band mag arcsec$^{-2}$ (57650 $L_{\odot}$ kpc$^{-2}$) isophote. 

The red ellipse in the middle panel of Figure \ref{fig:part_mock_circle} has a major axis equal to the best-fit $R_e$ and illustrates our method. The right panel shows a MLR map for this galaxy, which is effectively the ratio of the left panel to the middle panel when both are smoothed to the median smoothing length of the stellar particles.  We see that the MLR is far from uniform across the face of this galaxy -- a point that will be important for understanding our results in the next section.

The left panel of Figure \ref{fig:sersic} provides an illustration of the best-fit S\'ersic profile (black) of the same galaxy shown in Figure \ref{fig:part_mock_circle}.  We see that the single S\'ersic fit does a good job of matching the inner profile, though the true (g-band) luminosity profile (red) is more extended at large radii. The right panel shows the luminosity (red) and mass (blue) surface density profile of the same galaxy both smoothed to the median stellar smoothing length.  We see that the MLR increases from $\simeq 1$ near the center to $>2$ at large radii, as anticipated from the right panel of Figure \ref{fig:part_mock_circle}.

Table \ref{tab:sims} lists our best-fit values for $R_e$, $n$, and $b/a$ for each galaxy and galaxy orientation. The total $g$-band luminosity integrated from the S\'ersic profile $L_g$ and $g-r$ colour is also listed in Table \ref{tab:sims}.

To determine the ``mock'' stellar mass of each galaxy, $M_{\star}^{\rm mock}$, we follow common practice in much of the literature and assume a constant MLR based on the total integrated colour of the galaxy.~\footnote{Note that, as can be seen in the right panel of Figure \ref{fig:part_mock_circle}, the MLR is not constant across the face of the galaxy, suggesting that such an approach may fall short in reproducing the true underlying mass.} We specifically use colour-stellar mass-to-light ratio relation (CMLR) derived by \citet{Into2013NewDust} for low-mass galaxies using exponential star formation models:

\begin{equation}
  \log(M_{\star}^{\rm mock}/L_g) = 1.774 (g-r) - 0.783 \, .
  \label{eq:M/L}
\end{equation}

Here, $L_g$ is the total $g$-band luminosity integrated from the S\'ersic profile, and $g$ \& $r$ are the total magnitudes in the respective bands. Note that \cite{Into2013NewDust} argue that dust is a second order effect in the optical and do not include dust corrections in this CMLR.~\footnote{We concur with this assessment. For example, if we change our dust attenuation model from an SMC-inspired case to a Milky-Way inspired model, our results change only minimally.} Also, while the \cite{Into2013NewDust} relation assumes a Kroupa (2001) IMF rather than a Chabrier (2003) IMF, as we have done in our mock images, the two IMF assumptions yield nearly identical predictions for MLR of mono-age populations at fixed metallicity (see Figure 4, \cite{Bruzual2003Stellar2003}). We chose this particular CMLR relation because it was used in the ELVES sample, which dominates our observational comparison below \citep[][]{Carlsten2021StructuresShapes}. In Table \ref{tab:sims} we also report the implied ``true'' MLR of our galaxies ($M_{\star}^{\rm part}$/$L_{\rm image}$) where the luminosity is integrated from the full $g$-band image and not the S\'ersic profile.

\section{Results}
\label{sec:results}

\begin{figure*}
  \centering
  \includegraphics[width=.9\textwidth]{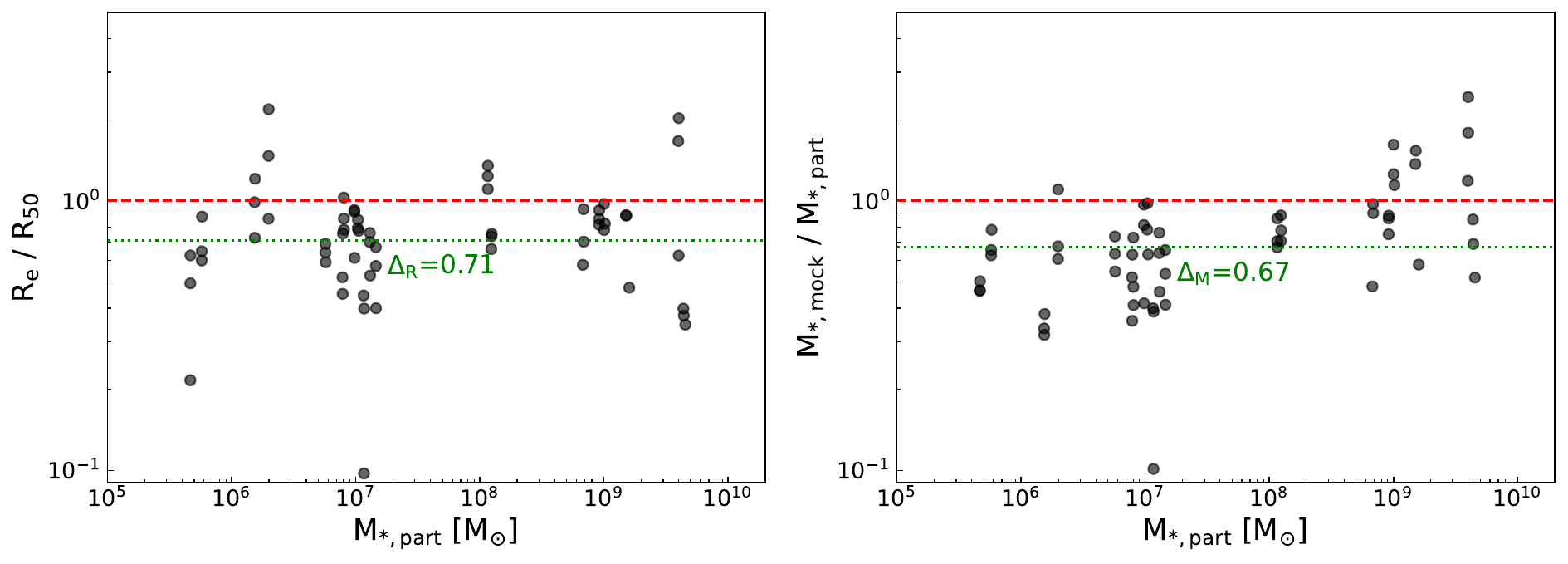}
  \caption{
  \textit{Left panel:} For a given simulated galaxy, we show the ratio of the half-mass radius to the effective radius with respect the galaxy's total stellar mass found through the particle method.
  \textit{Right panel:} Here we show the ratio between the total stellar mass found with the particle method and the total stellar mass found through the mock observation method versus the particle total stellar mass. The red dashed line shows the 1:1 for the Mock/Particle quantities and the green dotted line shows the average Mock/Particle quantities ($\Delta_R$, $\Delta_M$).
  }
  \label{fig:r50/re_Mi/Mm}
\end{figure*}

\subsection{Size-Mass Relation}
\label{sec:simvobs}

The left panel of Figure \ref{fig:mvr} shows the particle-based SMR of our simulated galaxies (black triangles) with the dotted black line being the best fit relation and the 1$\sigma$ scatter shown as the black shaded region. We compare to the results of two observational surveys: Galaxy And Mass Assembly (GAMA) \citep{Lange2016GalaxySpheroids} (in blue) and the Exploration of Local Volume Satellites Survey (ELVES) \citep{Carlsten2021StructuresShapes} (in red). We show the best fit relation and 1$\sigma$ for the GAMA late-type galaxy sample and the ELVES isolated galaxy sample found in the Updated Nearby Galaxy Catalog \citep{Karachentsev2013SUITESGALAXIES}.

In the right panel of Figure \ref{fig:mvr} we use our mock-derived sizes and stellar masses to make the same comparison. Simulated galaxies are shown as plus signs in this plot. Remarkably, both the best-fit relation (black dotted line) and increased scatter (shaded grey) match the observed trends much more closely than in the particle-based method.

Quantitatively, our particle-based measurements produce a power-law relation $R \propto M_\star^\alpha$ with $\alpha = 0.291$, which is quite similar to the slope of the ELVES sample ($\alpha = 0.296$). However, at fixed stellar mass, the particle-based sizes are $\approx$ 30\% larger than the observed relations suggest they should be. Moreover, the $1 \sigma$ scatter is $\approx$ 50\% smaller in the particle-based predictions than reported for the ELVES sample.  The mock-derived relations, on the other hand, continue to have a similar slope $\alpha = 0.289$, but now the predicted scatter and normalization are almost identical to those reported in the observational samples shown.

The shift in the relation from particle-based to mock-based is illustrated by the inset shown in the right panel of Figure \ref{fig:mvr}. This is a zoomed in version of the simulated SMR with both the particle method and the mock observation method show as a dotted and dashed line respectively. On average, at fixed stellar mass, the mock galaxy sizes are smaller by a factor of $0.71$ ($-$0.15 dec) than the particle-based sizes (green down arrow). Similarly, the mock-inferred stellar masses are smaller by a factor of $0.67$ ($-$0.17 dec) than the particle-inferred masses (green left arrow). Ultimately, the mock relation ends up having a similar slope but a lower normalisation than the particle-based relation. We dive into what is causing these shifts in the next session.

\subsection{Differences between the Particle-derived and Mock-derived Quantities}
\label{sec:partvmock}

Our goal in this section is to understand the differences in galaxy sizes and stellar masses that arise when comparing our particle method to our mock-observation method.

The left panel of Figure \ref{fig:r50/re_Mi/Mm} plots the ratio of mock half-light radius to particle-derived half-mass radius ($R_e/R_{50}$) for each galaxy orientation (black circles) as a function of particle-inferred stellar mass, $M_{\star}^{\rm part}$. While there are times when this ratio is larger than unity, on average we find $\langle R_e/R_{50} \rangle \equiv \Delta_R = 0.71$. Note that $R_e$ is typically {\em smaller} than $R_{50}$ even though $R_e$ is the major-axis effective radius while $R_{50}$ assumes circular symmetry.  Allowing for an elliptical estimate of $R_{50}$ would likely only make this ratio smaller. The origin of this difference is that the predicted g-band light tends to be more centrally concentrated than the mass, which is consistent with the MLR map shown in the right panel of Figure \ref{fig:part_mock_circle} and in the central panels of Figure \ref{fig:bulk1} in the Appendix. A positive MLR gradient is expected if recent star formation and more luminous stars tend to inhabit the center of the galaxy with less-massive, older stars in the outer regions. This is indeed the case for FIRE galaxies in this mass range, and has been shown to arise as a result of star-formation ``puffing'' \citep{El-Badry16,Graus19,Mercado21}.

Observationally, it is know that low-mass galaxies tend to have positive MLR gradients \citep[e.g.,][]{Tortora2011StellarMass} in qualitative agreement with our simulations. Higher mass galaxies, on the other hand, are observed to  have flat or negative gradients, depending on the galaxy type \citep{Tortora2010ColourMass,Tortora2011StellarMass}. Given this, we might expect $R_e/R_{50}$ to be greater than or equal to unity for more massive galaxies than we explore here.

The right panel of Figure \ref{fig:r50/re_Mi/Mm} shows the ratio of mock-derived and particle-derived stellar masses, M$_{\star}^{\rm mock}/$M$_{\star}^{\rm part}$, as a function of particle-derived stellar mass. There is a hint that the ratio increases as stellar mass increases, with the highest mass bin showing no clear systematic bias. However, for most of the mass range shown, the ratio is biased toward low values. Over the entire sample we find an average $\langle {\rm M}_{\star}^{\rm mock}/{\rm M}_{\star}^{\rm part} \rangle \equiv \Delta_M = 0.67$. 

\begin{figure}
  \centering
  \includegraphics[width=.45\textwidth]{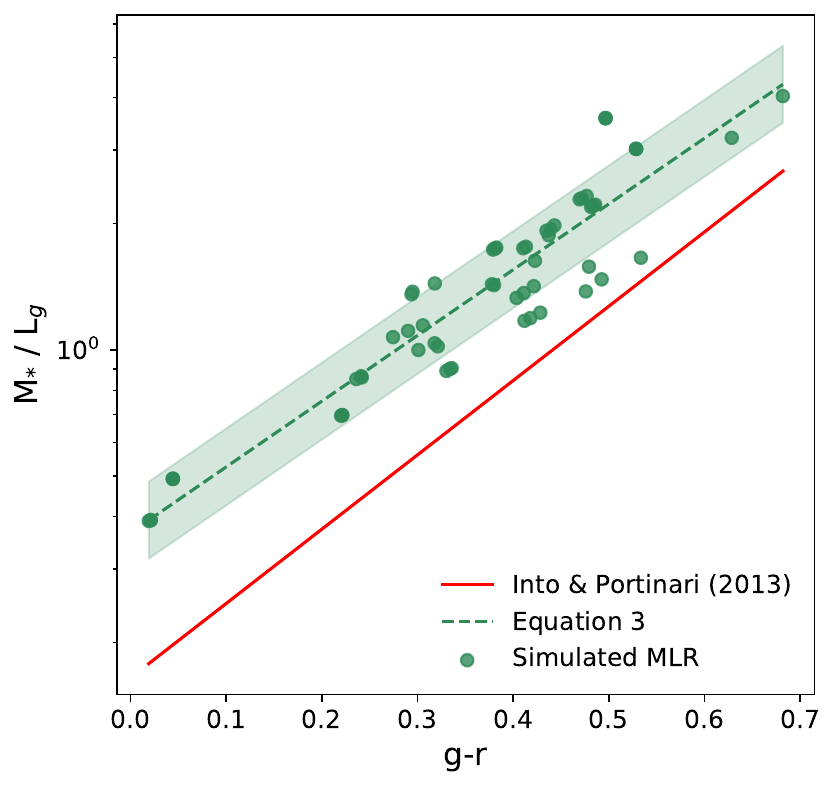}
  \caption{The colour-stellar mass-to-light ratio relation (CMLR) for our simulated galaxies (green points). The red line shows the CMLR from \citet{Into2013NewDust}, which we have adopted in our mock-derived properties in order to conform with assumptions used in the observational samples shown in Figure \ref{fig:mvr}. The green dashed line shows the best-fit to the simulations, given in Equation \ref{eq:M/L_sim}. The green shaded band shows the 1 $\sigma$ scatter. }
  \label{fig:g-r_ml}
\end{figure}

One important reason that our mock stellar masses are larger than those derived from particle counting is that the CMLR relation adopted to conform with observational assumptions (Equation \ref{eq:M/L}) is a poor match to the relation present in our simulations. The green points in Figure \ref{fig:g-r_ml} show the true CMLR of our simulated galaxies, derived using the total particle-derived stellar mass, together with mock luminosities and magnitudes of our galaxy images in the $g$ and $r$ band. We see that these points have systematically higher stellar MLRs at fixed colour than the relation from \citet{Into2013NewDust}, which we show as the red line. The dashed green line shows the best-fit CMLR for our simulated galaxies: 
 
\begin{equation}
  \log(M/L) = 1.570 (g-r) - 0.437 \, .
  \label{eq:M/L_sim}
\end{equation}
This can be compared to Equation \ref{eq:M/L}, which has a slightly steeper slope and lower normalisation.

The main difference seen in Figure \ref{fig:g-r_ml} is that the CMLR for our simulated galaxies gives higher MLR ratios than Equation \ref{eq:M/L}. One likely cause of this shift is that the models explored in \citet{Into2013NewDust} assume exponential star formation histories. Our simulated FIRE-2 galaxies have temporally extended, complex star formation histories \citep[e.g.][]{Fitts17,Garrison-Kimmel2019StarEnvironment,Flores-Vel2021}. Exponential histories, in contrast, will tend to have more recent star formation at fixed colour, biasing the mass estimates low. 

Others have found that the assumption of exponential star formation histories can bias mass estimates low compared to more generalized possibilities. For example, \citet{Li2022FlexibleRelationship}  explored the effect of a star formation history on the inferred CMLR of observed galaxies. They found that a non-parametric star formation history was the main parameter in their SED model that caused an increase in the MLR by 0.12 dex when compared to a simple SED model that assumed an exponentially-declining star formation history.  The qualitative difference here is in line with the shifts we see. Moreover, these authors focus on higher redshifts, when the timescales are more compressed.  If anything, we expect the differences between exponential and extended star formation to be even larger at low redshift, closer to the 0.25 dex we see.

A second driver relates to the MLR gradient discussed above. Our mock estimates assume a spatially constant MLR, however, in reality, our galaxies have older, dimmer populations at large radii that contain an over-abundance of mass than accounted for in a mass-follows-light assumption. If galaxies in the real universe behave more like our FIRE-2 galaxies, then Equation \ref{eq:M/L_sim} would be a better choice for estimating their stellar masses from $g-r$ colour.

\section{Conclusion}
\label{sec:conclusions}

We present a study of the size versus stellar-mass relation (SMR) for low-mass galaxies in the FIRE-2 simulations and explore the differences that arise when adopting observationally-motivated ``mock'' measurements compared to a more traditional ``theorist'' approach that utilises particle-counting directly. Our mock observation method employs S\'ersic profile fits to simulated images. We provide fit parameters, including axis ratios and S\'ersic indices in Table \ref{tab:sims}. Our primary conclusions are the following:

\begin{enumerate}

  \item While our particle-based measurements of simulated galaxy sizes and masses produce an SMR that is offset high and with less scatter at fixed stellar mass than published relations from ELVES and GAMA (left panel, Figure \ref{fig:mvr}), the same relation derived from our mock observations reproduces the normalisation, slope, and scatter in the observed relation remarkably well (right panel, Figure \ref{fig:mvr}).

  \item Our mock-galaxy half-light radii are $\sim 30\%$ smaller than the half-mass radii inferred from particle-counting directly (left panel, Figure \ref{fig:r50/re_Mi/Mm}). This is primarily due to the fact that our galaxies have mass-to-light ratio gradients (Figures \ref{fig:part_mock_circle}, \ref{fig:sersic}, and \ref{fig:bulk1}), with older (dimmer) stellar populations at larger radii. 

  \item The particle-based stellar masses of our galaxies are $\sim 30\%$ larger than our mock-inferred stellar masses (right panel, Figure \ref{fig:r50/re_Mi/Mm}). This is because our mock stellar masses rely on a colour-stellar mass-to-light ratio relation (CMLR) that under-predicts the true value (Figure \ref{fig:g-r_ml}). Our mock CMLR is the same as that assumed by the ELVES group \citep{Into2013NewDust} in deriving their published SMR. This assumption under-predicts the stellar masses of our galaxies, likely because our galaxies have complex, temporally extended star-formation histories that are not well represented by exponential functions. Equation \ref{eq:M/L_sim} provides an alternative CMLR that matches our predictions for low-mass galaxies. 

\end{enumerate}

This work highlights the nuances associated with testing simulations against observations as well as some potential pitfalls in interpreting the physical meaning of observational results. Our simulations are able to reproduce published size-mass relations of low-mass galaxies if we perform mock observations that mimic assumptions adopted in those papers. However, the true relationship between stellar mass and size present in the simulation is different. If our simulations are accurate representations of the real universe, then the relationship between galaxy size and stellar mass is even tighter for low-mass galaxies than might be expected from taking the published observational results at face value.

\section*{Acknowledgements}
CK is supported by a National Science Foundation (NSF) Graduate Research Fellowship Program under grant DGE-1839285. JSB is supported by NSF grant AST-1910965 and NASA grant 80NSSC22K0827. JM is supported by the Hirsch Foundation. FJM is supported by the NSF MSP-Ascend Award AST-2316748. Support for PFH was provided by NSF Research Grants 1911233, 20009234, 2108318, NSF CAREER grant 1455342, NASA grants 80NSSC18K0562, HST-AR-15800. RKC is funded by support for program \#02321, provided by NASA through a grant from the Space Telescope Science Institute, which is operated by the Association of Universities for Research in Astronomy, Inc., under NASA contract NAS 5-03127. JAB acknowledges support from NSF grant AST-2107993. Numerical calculations were run on the Caltech compute cluster allocations AST21010 and AST20016 supported by the NSF and TACC, and NASA HEC SMD-16-7592. 
We thank the anonymous referee for helpful comments that improved the discussions and quality of the paper.

\section*{Data Availability}
The data supporting the plots within this article are available on reasonable request to the corresponding author. A public version of the {\small GIZMO} code is available at \url{http://www.tapir.caltech.edu/~phopkins/Site/GIZMO.html}. Additional data including simulation snapshots, initial conditions, and derived data products are available at \url{https://fire.northwestern.edu/data/}. Some of the publicly available software packages used to analyze these data are available at: \url{https://bitbucket.org/awetzel/gizmo\_analysis}, and \url{https://bitbucket.org/awetzel/utilities}.




\bibliographystyle{mnras}
\bibliography{bib} 




\appendix
\section{Example galaxy images and profiles}
\label{sec:appendix}

\begin{figure*}
  \centering
\includegraphics[width=.49\textwidth]{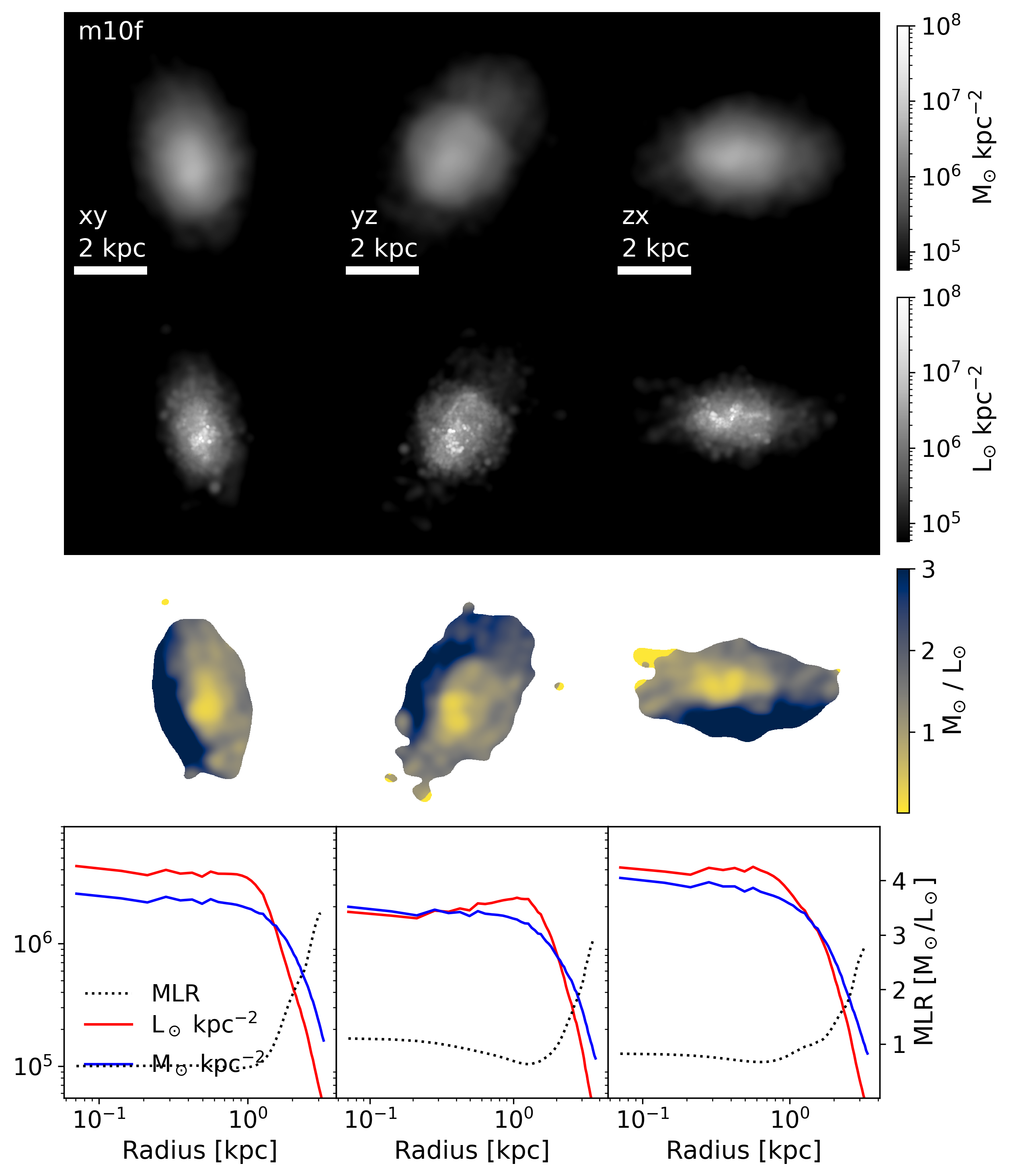}
\includegraphics[width=.49\textwidth]{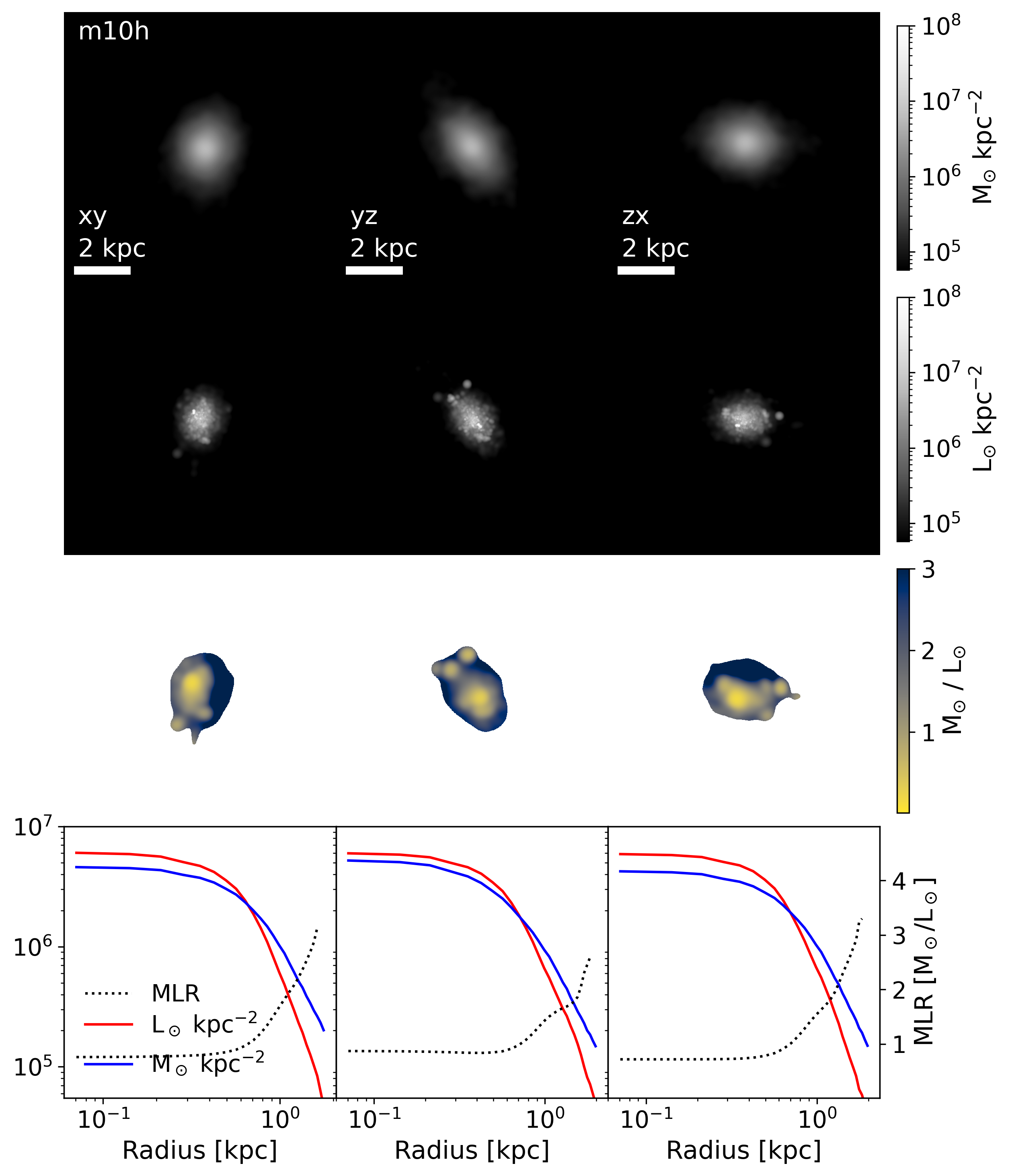}
\includegraphics[width=.49\textwidth]{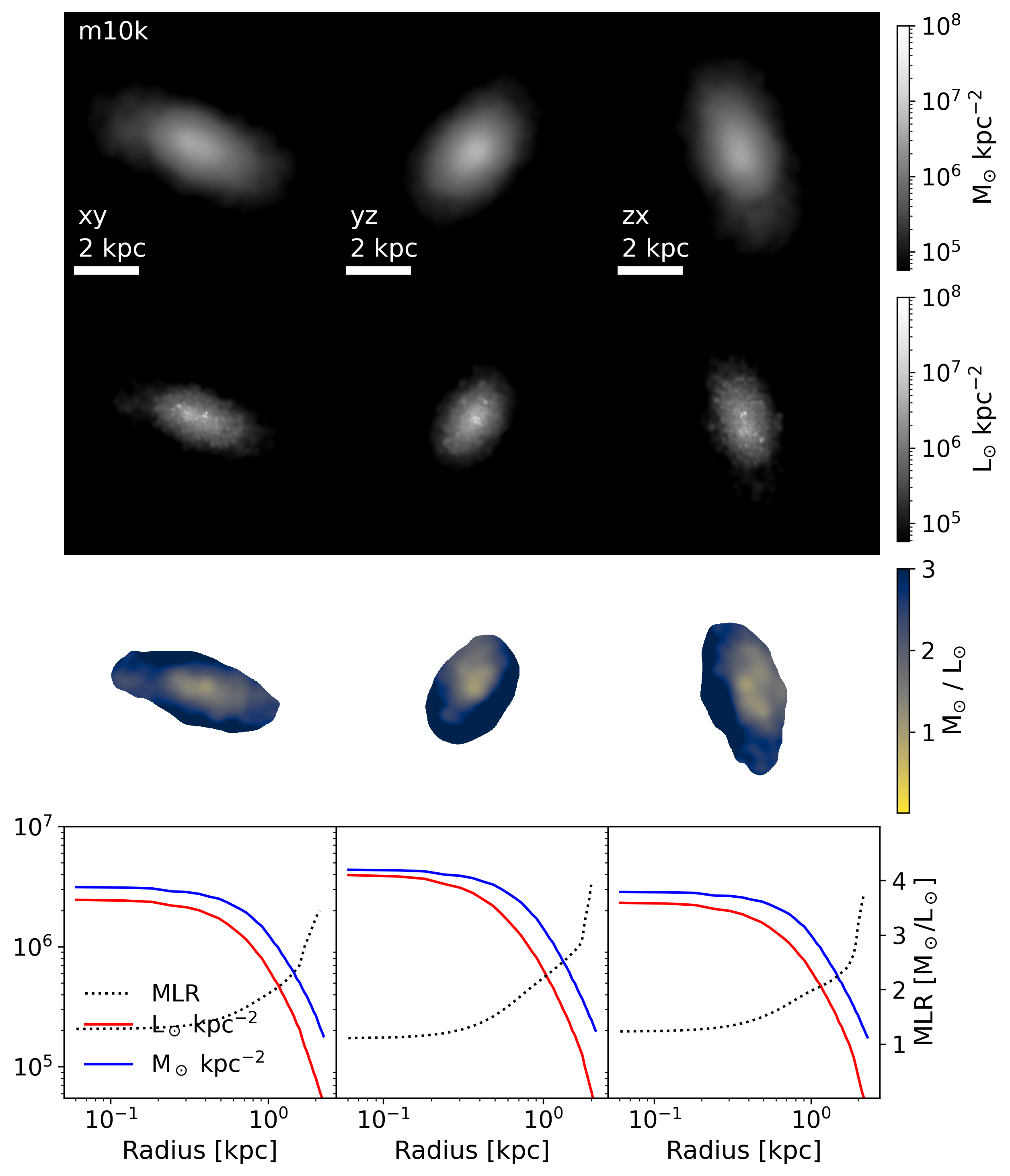}
\includegraphics[width=.49\textwidth]{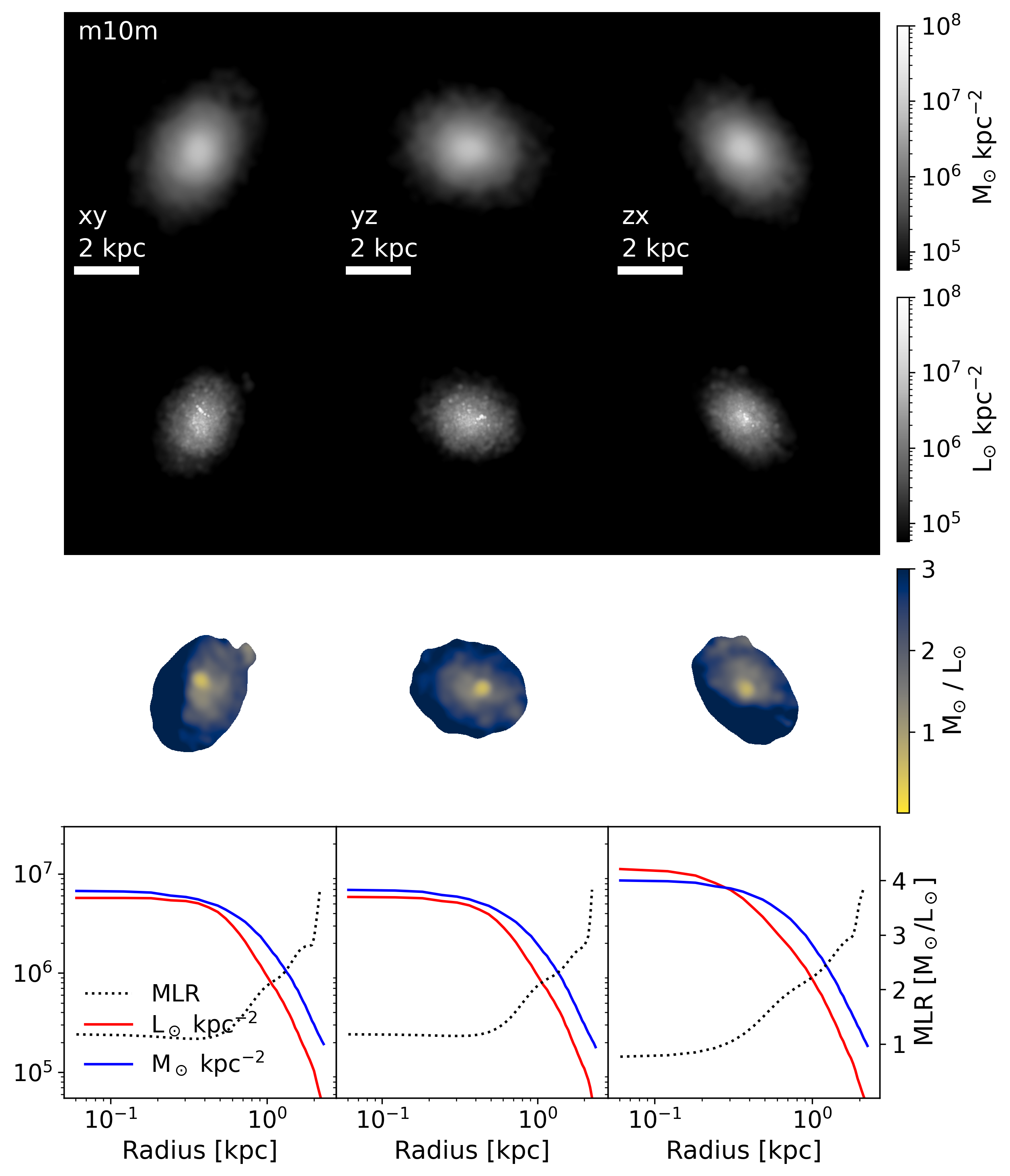}
  \caption{Four example galaxies -- m10f (top left), m10h (top right), m10k (bottom left), and m10m (bottom right) -- each shown in projection along three different axes (columns). The top row and second row in each set show surface mass density and g-band luminosity density maps along the projections. The third row in each set shows implied MLR maps. The bottom row shows surface mass density (blue) and luminosity density (red) profiles as a function of semi-major axis radius, as in Figure \ref{fig:sersic}. The dotted lines are the implied MLR profiles (right axis).} 
  \label{fig:bulk1}
\end{figure*}

\begin{figure*}
  \centering
\includegraphics[width=.49\textwidth]{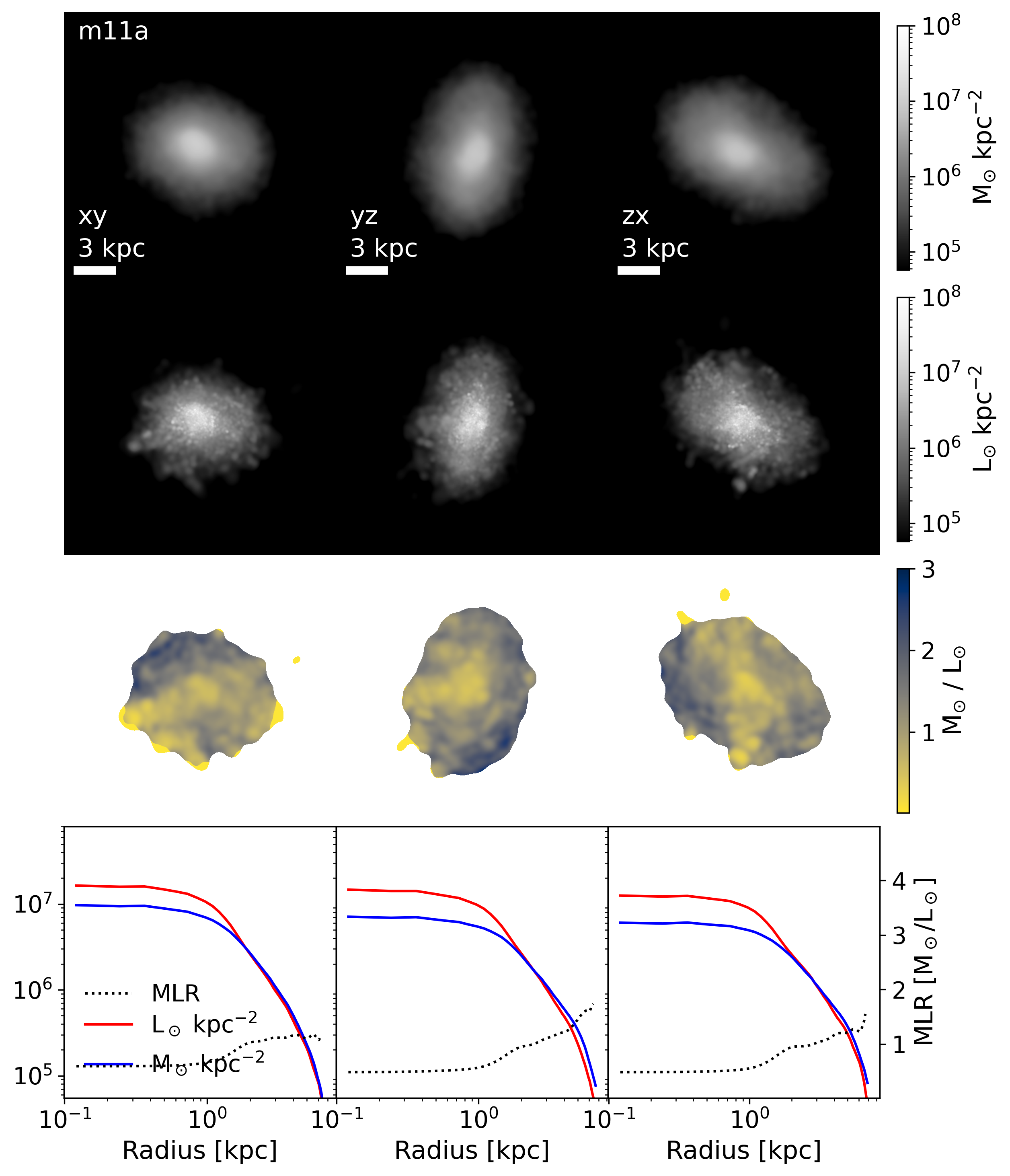}
\includegraphics[width=.49\textwidth]{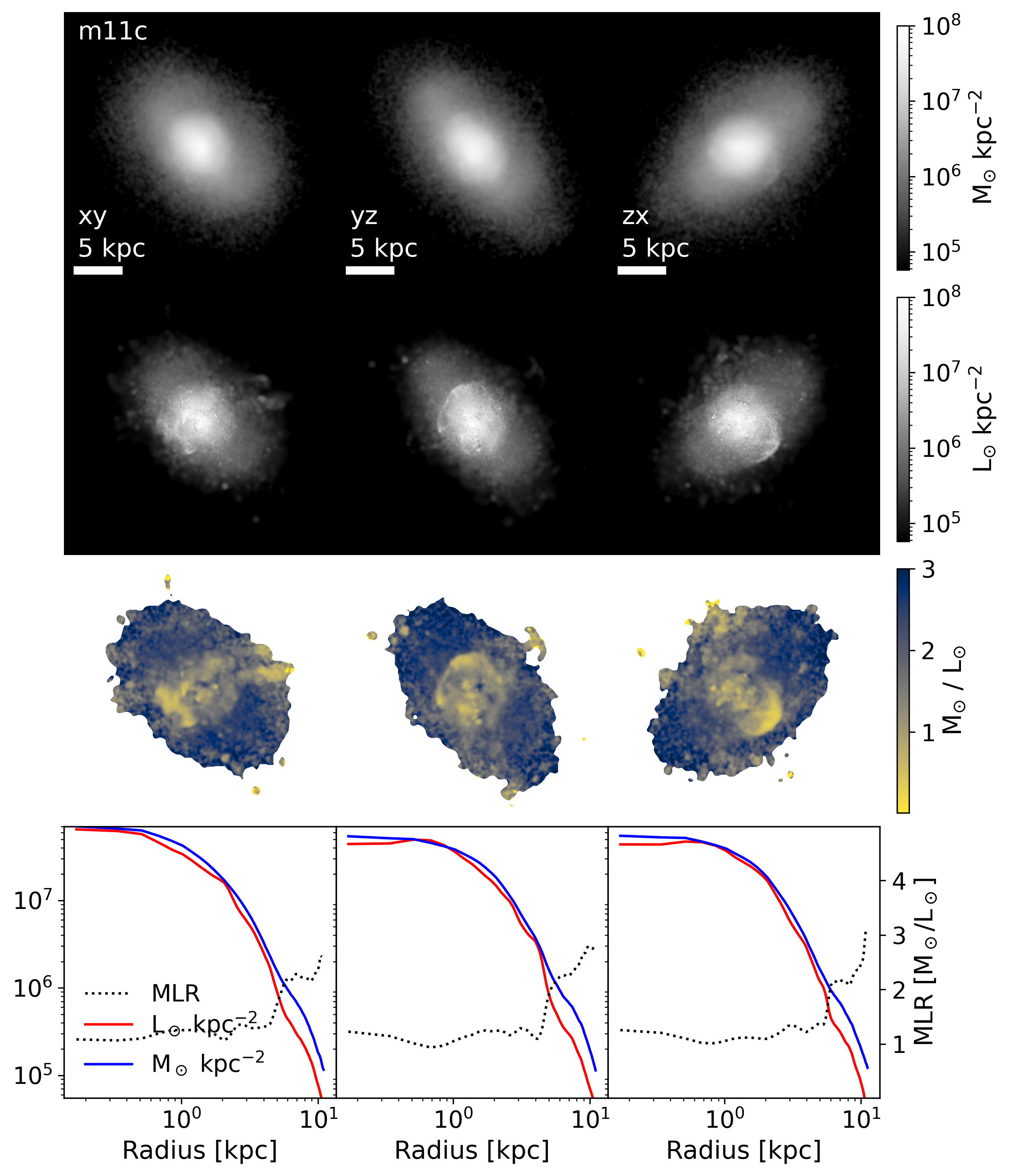}
\includegraphics[width=.49\textwidth]{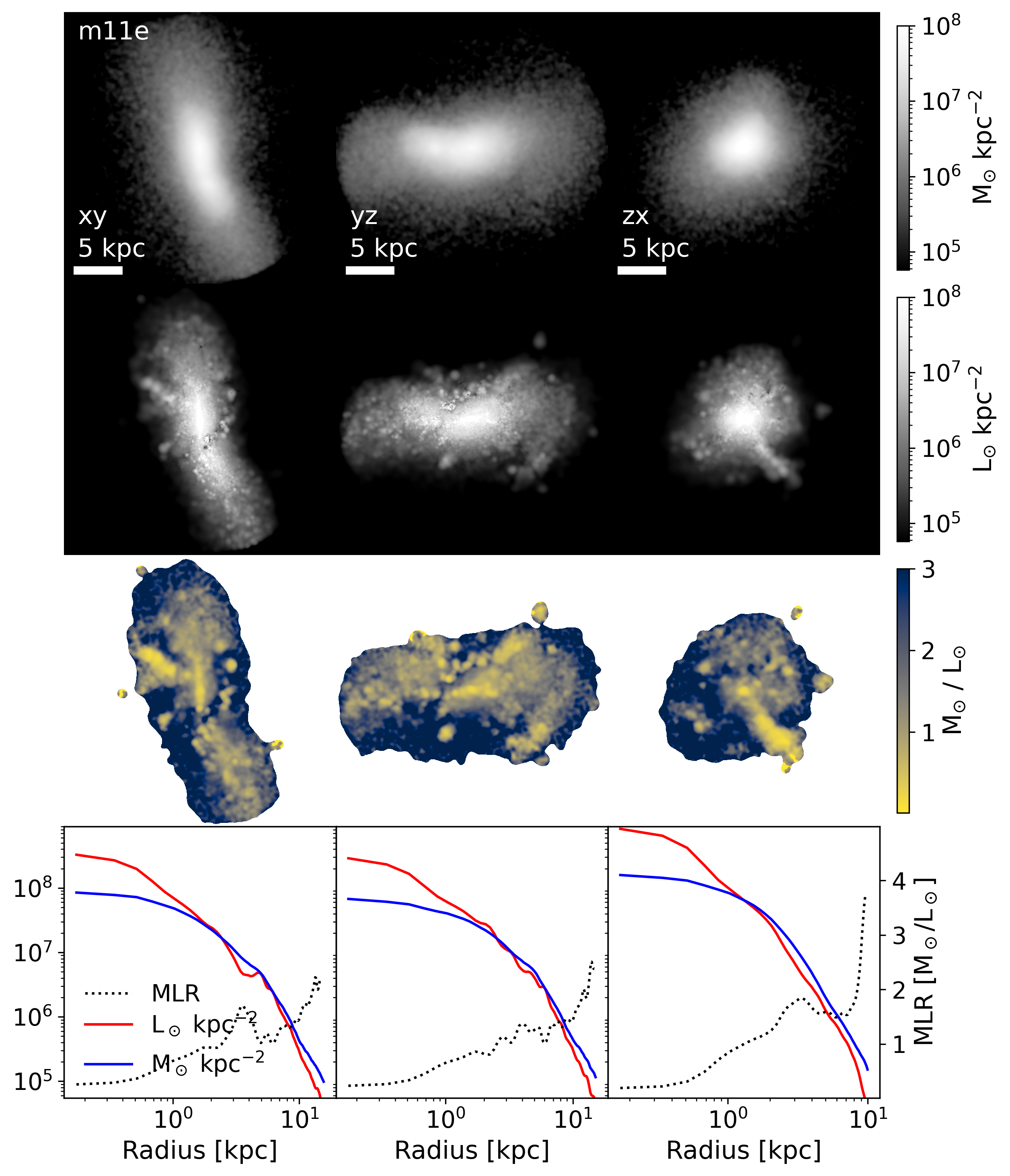}
\includegraphics[width=.49\textwidth]{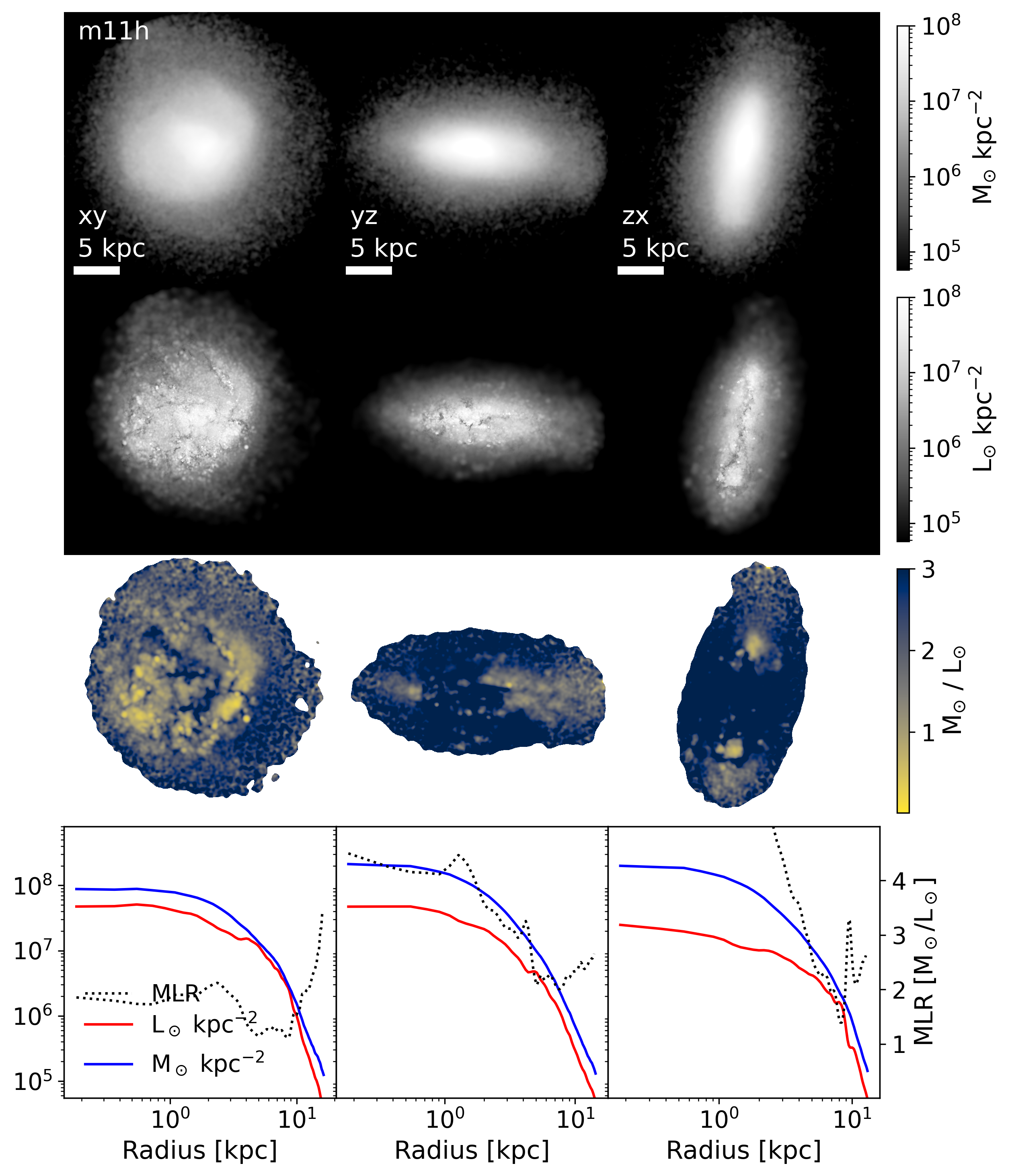}
  \caption{Same as Figure \ref{fig:bulk1} except for galaxies m11a, m11c, m11e, and m11h.}
  \label{fig:bulk2}
\end{figure*}

Figures \ref{fig:bulk1} and \ref{fig:bulk2} provide example mass surface density, luminosity surface density, and MLR maps in three different orientations for a representative sample of galaxies that span the mass and morphology of our survey. Also shown are mass surface density profiles, luminosity surface density profiles, and MLR profiles for each orientation. These panels are similar to those shown in Figures \ref{fig:part_mock_circle} and \ref{fig:sersic} shown in the main text.

\bsp	
\label{lastpage}

\end{document}